\documentclass{aastex}
\usepackage{graphicx}
\usepackage{emulateapj5}
\usepackage{times}
\usepackage{color} 



\def\chandra    {{\em Chandra}\/}

\def\suzaku     {{\em Suzaku}\/}

\def\ga         {{ESO~137-001}\/}
\def\gab        {{ESO~137-002}\/}

\usepackage{mathptmx}

\begin{document}
\renewcommand*{\thefootnote}{\fnsymbol{footnote}}

\title{The Narrow X-ray Tail and Double H$\alpha$ Tails of \gab\ in Abell 3627\footnotemark}

\footnotetext{Based on observations made with the \chandra\ X-ray Observatory and the Southern Observatory for Astrophysical Research (SOAR) telescope.}

\author{
B.\ Zhang$^{\!}$\altaffilmark{1,2},
M.\ Sun$^{\!}$\altaffilmark{3},
L.\ Ji$^{\!}$\altaffilmark{4},
C.\ Sarazin$^{\!}$\altaffilmark{5},
X.\ B.\ Lin$^{\!}$\altaffilmark{1,2},
P.\ E.\ J.\ Nulsen$^{\!}$\altaffilmark{6},
E.\ Roediger$^{\!}$\altaffilmark{7},
M.\ Donahue$^{\!}$\altaffilmark{8},
W.\ Forman$^{\!}$\altaffilmark{6},
C.\ Jones$^{\!}$\altaffilmark{6},
G.\ M.\ Voit$^{\!}$\altaffilmark{8},
and X.\ Kong$^{\!}$\altaffilmark{1,2}
}

\smallskip

\affil{\scriptsize 1) Center for Astrophysics, University of Science and Technology of China, Hefei, Anhui, 230026, China}
\affil{\scriptsize 2) Key Laboratory for Research in Galaxies and Cosmology, USTC, Chinese Academy of Sciences, China}
\affil{\scriptsize 3) Department of Physics, University of Alabama in Huntsville, Huntsville, AL 35899, USA}
\affil{\scriptsize 4) Purple Mountain Observatory, Chinese Academy of Sciences, Nanjing, Jiangsu, 210008, China}
\affil{\scriptsize 5) Department of Astronomy, University of Virginia, P.O. Box 400325, Charlottesville, VA 22904-4325, USA}
\affil{\scriptsize 6) Harvard-Smithsonian Center for Astrophysics, 60 Garden St., Cambridge, MA 02138, USA}
\affil{\scriptsize 7) Germany Hamburger Sternwarte, Universit\"at Hamburg, Gojensbergsweg 112, D-21029 Hamburg, Germany}
\affil{\scriptsize 8) Department of Physics and Astronomy, Michigan State University, East Lansing, MI 48824, USA}
\email{mingsun.cluster@gmail.com, xkong@ustc.edu.cn}
\shorttitle{Tails of \gab }
\shortauthors{Zhang et al.}

\begin{abstract}

We present the analysis of a deep \chandra\ observation of a \textbf{$\sim2L_*$} late-type galaxy, \gab , in the closest rich cluster A3627. The \chandra\ data reveal a long ($\gtrsim$ 40 kpc) and narrow tail with a nearly constant width ($\sim3$ kpc) to the southeast of the galaxy, and a leading edge $\sim1.5$ kpc from the galaxy center on the upstream side of the tail. The tail is most likely caused by the nearly edge-on stripping of \gab 's interstellar medium (ISM) by ram pressure, compared to the nearly face-on stripping of \ga\ discussed in our previous work. Spectral analysis of individual regions along the tail shows that the gas throughout it has a rather constant temperature, $\sim1$ keV, very close to the temperature of the tails of \ga , if the same atomic database is used. The derived gas abundance is low ($\sim0.2$ solar with the single-$kT$ model), an indication of the multiphase nature of the gas in the tail. The mass of the X-ray tail is only a small fraction ($<5\%$) of the initial ISM mass of the galaxy, suggesting that the stripping is most likely at an early stage. However, with any of the single-$kT$, double-$kT$ and multi-$kT$ models we tried, the tail is always ``over-pressured" relative to the surrounding intracluster medium (ICM), which could be due to the uncertainties in the abundance, thermal vs. non-thermal X-ray emission, or magnetic support in the ICM. The H$\alpha$ data from the Southern Observatory for Astrophysical Research (SOAR) show a $\sim21$ kpc tail spatially coincident with the X-ray tail, as well as a secondary tail ($\sim12$ kpc long) to the east of the main tail diverging at an angle of $\sim23^\circ$ and starting at a distance of $\sim7.5$ kpc from the nucleus. At the position of the secondary H$\alpha$ tail, the X-ray emission is also enhanced at the $\sim2\sigma$ level. We compare the tails of \ga\ and \gab , and also compare the tails to simulations. Both the similarities and differences of the tails pose challenges to the simulations. Several implications are briefly discussed. 

\end{abstract}


\section{Introduction}

It has long been known that galaxies occur in different environments and many galactic properties depend on environment, e.g., Hubble type, color, and star formation rate.  (See the extensive review by Boselli \& Gavazzi 2006 and references therein.)  This implies that some processes related to the environment must be at work. Interaction with the ICM, one of these processes, is likely to be dominant in clusters in the present universe (Boselli \& Gavazzi 2006). Ram pressure, the drag force acting on the gas in a galaxy when it moves in the ICM, can strip the cold ISM down to the stripping radius beyond which the ram pressure exceeds the gravitational restoring force.  Thus, ram pressure stripping can decrease the gas content, quench the global star formation rate (SFR), and lead to a morphological transformation of the galaxy (e.g., Gunn \& Gott 1972; Vollmer et al.\ 2001b; Crowl \& Kenney 2008; Quilis et al.\ 2000).   

On the other hand, once the cold ISM has been stripped out of the galaxy, it mixes with the hot ICM and eventually becomes part of it. Due to star formation in the galaxy, the ISM initially should be metal rich, and stripping of this gas will enrich the ICM. Stripping of the ISM can also contribute to the clumping of the ICM, which will induce significant biases in observed ICM profiles based on X-ray measurements, further affecting the use of clusters of galaxies as precise probes of cosmology (Nagai \& Lau 2011). The stripped tails, with the rich range of microphysical processes occurring within them and at their interfaces with the ICM (e.g., Cowie \& McKee 1977; Nulsen 1982; Roediger \& Hensler 2005; Kapferer et al.\ 2009; Tonnesen \& Bryan 2011), provide an excellent environment to study the baryonic physics which is crucial to understand structure formation and evolution.  

Since the first analytical discussion by Gunn \& Gott (1972), much evidence for ram pressure stripping has been found, either by the deficit of H I gas (e.g., Giovanelli \& Haynes 1985; Cayatte et al.\ 1994; Vollmer et al.\ 2001b) or tails of stripped gas (e.g., in H I: Kenney et al.\ 2004; Oosterloo \& van Gorkom 2005; Chung et al.\ 2007, 2009; in H$\alpha$: Gavazzi et al.\ 2001a; Sun et al.\ 2007, hereafter S07; Yagi et al.\ 2007; Yoshida et al.\ 2008; Smith et al.\ 2010; Yagi et al.\ 2010; Fossati et al.\ 2012; in X-rays: Finoguenov et al.\ 2004; Wang et al.\ 2004; Rasmussen et al.\ 2006; Sun et al. 2006; Sun et al. 2010, hereafter S10; We$\dot{\rm z}$gowiec et al.\ 2011). Meanwhile, much theoretical work has been done with more realistic modeling of the galaxy structure and a fuller treatment of the processes, e.g., multiphase ISM (Tonnesen \& Bryan 2009), varying winds (Schulz \& Struck 2001; Roediger \& Br$\ddot{\rm u}$ggen 2008b), star formation (Kronberger et al.\ 2008; Fujita \& Nagashima 1999), and magnetic fields (Ruszkowski et al.\ 2012). In our previous work (S10), we presented the discovery of the spectacular X-ray tails attached to a late-type galaxy, \ga , in the massive cluster A3627. The tails are long ($\sim80$ kpc) and narrow ($<8$ kpc) with nearly constant widths and temperatures ($\sim0.8$ keV). Star formation is active in the stripped gas, with many H II regions and young star clusters (S07; S10). The most remarkable feature is the double tails extending well outside of the tidally truncated dark matter halo, making \ga\ the first late-type galaxy with long double X-ray tails (S10). S10 also presented the discovery of an X-ray tail and a H$\alpha$ tail associated with another late-type galaxy in A3627, \gab , which are the focus of this paper.
 
A3627 is the closest massive cluster ($z=0.0163$, $\sigma_{\rm rad} = 925$ km s$^{-1}$ and $kT=6$ keV), rivaling the prominent Perseus and Coma clusters in mass and galaxy content. It is located behind the Milky Way ($b\sim-7^\circ$) near the center of the Great Attractor (Kraan-Korteweg et al.\ 1996; Woudt et al.\ 2004, 2008). \gab\ ($z=0.0191$) is a more massive and redder galaxy than \ga\ (see Table 1 for the basic parameters of the two galaxies), and is viewed edge-on with a dust lane across its bulge. \gab\ also hosts a Seyfert-2-like nucleus (S10). S10 presented the properties of \gab 's X-ray tail from a 10$'$.8 off-axis \chandra\ observation with poor angular resolution. The galaxy is located near the edge of the CCD in that observation, which makes the subtraction of background uncertain. Now with a new, on-axis deep \chandra\ observation, the properties of \gab 's X-ray tail can be much better constrained.

This paper is structured as follows: The observations and data reduction are presented in Section 2. Section 3 describes the spatial structure and spectral properties of \gab 's tails, while the surrounding ICM is studied in Section 4. We discuss some properties, particularly the pressure, of \gab 's tails and compare them with other observations, mainly with \ga 's tails, and then with simulations in Section 5. Finally the main results of our paper are summarized in Section 6. We adopt $H_{0}=71$ km s$^{-1}$ Mpc$^{-1}$, $\Omega$$_{\rm M}=0.27$, and $\Omega_{\rm \Lambda}=0.73$, which yield a luminosity distance of 69.7 Mpc to the galaxy and the scaling 1$''=0.327$ kpc.

\section{Observations and data reduction}

The \chandra\ observation of \gab\ was conducted with the Advanced CCD Imaging Spectrometer (ACIS) on 2011 January 10-11 (ObsID 12950, PI: Sun). The observation was telemetered in VFAINT mode. Standard \chandra\ data analysis was performed with CIAO v4.4.1 and calibration database v4.5.0. We started from the Level 1 events lists and applied the CIAO tool \texttt{chandra\_repro} to obtain the new events files. The light curve from source-free regions of the S3 chip was examined. No background flares were found. The effective exposure is 89.56 ks for the S3 chip. In the spectral analysis, a lower energy cutoff of 0.5 keV was set to minimize calibration uncertainties. The absorption was fixed at the  Galactic value toward the direction of \gab , 1.73$\times10^{21}$ cm$^{-2}$ from the Leiden/Argentine/Bonn H I survey (Kalberla et al.\ 2005). The spectral fits were performed with XSPEC. AtomDB (the atomic database adopted by XSPEC) v2.0 released in 2011 includes significant changes of the Fe L-shell data, which affect the spectral fits for $kT<2$ keV plasma (Foster et al.\ 2012). However, work published prior to 2011 used AtomDB v1.3.1 or older versions, including S10. Therefore, the results with both AtomDB v2.0.1 and AtomDB v1.3.1 are listed. The C-statistic in XSPEC was adopted and uncertainties quoted are $1\sigma$.

The details of the SOAR observations and data reduction were presented in Section 6.1 of S10 (see also S07). The interested reader can refer to that section for more details.

\section{The Tails of \gab\ }

\subsection{Spatial Structure}

The $0.5-2.0$ keV \chandra\ count image is shown in panel \textbf{a} of Figure 1. Point sources were removed and the point source regions were filled with the surrounding background. The image was smoothed by a Gaussian kernel of 3$''$.94 to enhance the low surface brightness features. Neither background subtraction nor exposure correction was applied, so as to show the raw data in a minimally processed way. A long tail towards the southeast is clearly seen from the image. The background subtracted and exposure corrected image in the $0.5-2.0$ keV band is in panel \textbf{d} of Figure 1, with the main features marked. At the upstream side of the tail, a sharp, flattened discontinuity (edge) $\sim1.5$ kpc from the galaxy center (taken as the location of the X-ray AGN) is clearly visible, which is also shown in the surface brightness profile along the tail in Figure 2. The tail is swollen at the base, with a linear scale of $\sim5$ kpc for the widest part. Then the tail narrows down and changes direction, extending straight for $\sim5$ kpc. Afterwards it changes direction again and stays straight for another $\sim30$ kpc before curving in the direction to the cluster center and fading into the surrounding ICM. The swollen section and two straight parts of the tail, with a total length of $\sim40$ kpc in projection, are significantly above the local background, while the curved part is only a marginal feature (Figure 2). The tail is narrow, with a full width at half maximum (FWHM) of $2.9\pm 0.2$ kpc obtained for a Gaussian fit to the surface brightness profile across the tail (Figure 2). Furthermore, the width along the tail does not increase (i.e., it does not flare). This is in disagreement with simulations which do not include cooling (e.g., Roediger \& Br$\ddot{\rm u}$ggen 2008b; Tonnesen \& Bryan 2010). On the other hand, simulations with cooling produce significantly narrower and less flared tails (e.g., Tonnesen \& Bryan 2011), in better agreement with our observations.

Figure 1 also shows the optical images from SOAR: the net H$\alpha$ images are in panels \textbf{b} and \textbf{e}, while the optical continuum images are in panels \textbf{c} and \textbf{f}. The net H$\alpha$ image in panel \textbf{b} was smoothed with a 1$''$.5 Gaussian while that in panel \textbf{e} was smoothed with a 0$''$.6 Gaussian. The \chandra\ $0.5-2.0$ keV contours in red were overlaid on the images in panels \textbf{e} and \textbf{f}. Panels \textbf{b}, \textbf{c}, \textbf{e}, and \textbf{f} all have the same size. Because the cluster is very close to the Galactic plane, there are many foreground stars, as revealed in the optical continuum image. With the stars masked, the net H$\alpha$ image shows a tail extending to $\sim21$ kpc from the nucleus which spatially coincides with the X-ray tail. The X-ray leading edge also corresponds to the H$\alpha$ edge (Figure 3). These two facts imply that the X-ray tail and H$\alpha$ tail are physically one tail detected in different bands. 

A striking feature revealed by the H$\alpha$ image is the presence of a secondary tail to the southeast of the main tail. The secondary tail bifurcates at a distance of $\sim7.5$ kpc from the nucleus and diverges by $\sim23^\circ$ from the main tail, extending $\sim12$ kpc before becoming too faint to be observed. At the location of the H$\alpha$ secondary tail in the \chandra\ image, the X-ray emission is enhanced at the $\sim2\sigma$ level. A longer observation would be required to confirm the possible X-ray double tails.

\subsection{X-ray Spectral Properties}

We measured the gas temperatures in four regions along the X-ray tail, as shown in Figure 4. After excluding the central small circle (3$''$ in radius, centered on the X-ray AGN) from the green circle, we define the remainder as the ``head'' of the tail. The ``head" corresponds to the swollen part while the other three green boxes cover the two straight parts. Emission from the curved part is so weak that we decided not to measure the temperature in this region. The background is from local source-free regions. We fitted the spectra with the single-$kT$ (single APEC) model with absorption fixed at the Galactic value. The results are shown in the upper right of Figure 4. Mixing of the stripped ISM with the ICM naturally leads to a multiphase distribution of gas in the tail, especially when turbulence exists. As the cluster background is quite high and the regions for extracting spectra are small, we chose to keep the model simple. From Figure 4 (upper right panel), the temperature variations are small (within $1.4\sigma$) so the whole tail has a nearly constant temperature, which is similar to the X-ray tails of \ga . The global tail region, defined as the three green boxes (not including the ``head'') on the left of Figure 4, has a temperature of 0.95$^{+0.04}_{-0.05}$ keV with the single-$kT$ fit. However, the single-$kT$ fit yields a very low abundance (0.18$^{+0.09}_{-0.05}$ $Z_{\odot}$), which is a reflection of the multiphase nature of the gas \footnote { When a low resolution spectrum consisting of two or more $kT$ components (with the average temperature of $\sim1$ keV) is fitted with the single-$kT$ model, the Fe abundance will be seriously biased low (e.g., Buote 2000a, 2000b). For $0.5-1.5$ keV plasma, the temperature is determined by the centroid of the Fe-L hump. Two or more $kT$ components excite Fe-L lines over a wider energy range than the single-$kT$ model predicts, which means that the Fe-L hump is broader than the single-$kT$ model predicts. In order for the single-$kT$ model to better fit the spectrum, one has to reduce the size of the peak of the single-$kT$ model, i.e., the Fe abundance, as well as to increase the bremsstrahlung continuum (Buote 2000b; B$\ddot{\rm o}$hringer \& Werner 2010). As a result, the abundance will be biased low, since the abundance for $\sim1$ keV gas in APEC is essentially determined by the Fe abundance. Studies of cool-core clusters and groups have shown that the abundance increases when a double-$kT$ model or more complicated differential emission models are used (e.g., Buote et al.\ 2003; Werner et al.\ 2006).}.

Since the gas in the tail is multiphase in nature, we also tried double-$kT$ (double APEC) and multi-$kT$ (CEMEKL) models for spectral fitting. The results, as well as those with the single-$kT$ model, are shown in Table 2, in which we also list the fitting results for the ``head'' and the nucleus, defined as the small circle excluded from the ``head''. To compare with the X-ray tails of \ga\ in our previous work (S10), we also list the fitting results with AtomDB v1.3.1, which was used in that work. Figure 5 shows the spectral fit with a double-$kT$ model. The single-$kT$ fit gives a temperature of 0.95$^{+0.04}_{-0.05}$ keV, which can be compared with the S10 result, $1.98^{+0.96}_{-0.56}$ keV. The new data allow a better constraint on the local background, which results in a more accurate and better constrained temperature measurement.

The unfolded spectrum (i.e., deconvolved with the instruments' response under a given model) of the nucleus is displayed in Figure 6. A redshifted iron K$\alpha$ fluorescence line at $\sim6.3$ keV is present, with an equivalent width of 0.16$^{+0.06}_{-0.04}$ keV. The nucleus is luminous, with a bolometric luminosity of 1.2$\times10^{43}$ erg s$^{-1}$ (see Table 2). A bolometric correction factor of 15 is adopted from Vasudevan \& Fabian (2007). The high column density, $\sim2\times10^{23}$ cm$^{-2}$, confirms that \gab\ hosts an obscured AGN. Nishino et al.\ (2012) report the detection of excess hard X-ray emission from the center of A3627 with \suzaku. They checked the hard X-ray flux from a Seyfert-1 galaxy, IGR J16119-6036 (WKK 6092), and concluded that this flux is $4-10$ times too small to explain the observed hard X-ray emission as combination of the thermal emission from the ICM and a power-law from the Seyfert-1 galaxy. Since \gab\ hosts a luminous AGN, we compare its hard X-ray ($20-40$ keV) flux by extrapolating the \chandra\ spectrum to that from IGR J16119-6036, and find that it is $\sim3-6$ times smaller. Therefore, the hard X-rays from both IGR J16119-6036 and \gab\ are insufficient to fully account for the hard excess reported by Nishino et al.\ (2012).

\section{The Surrounding ICM}

The spectral properties of the ICM surrounding the tail were also examined. As the cluster emission and the X-ray background fill the whole chip, we used the stowed background, normalized to our observation in the $9.5-12$ keV band,  to subtract the non-X-ray background.  Following Sun (2009), we modeled the X-ray background with three components: a power-law with a photon index of 1.5 and an APEC thermal model, both with Galactic absorption, and an unabsorbed APEC thermal model with a fixed temperature of 0.1 keV. Both APEC thermal models had zero redshift and solar abundances. Another absorbed APEC thermal model with Galactic absorption was added to describe the cluster emission. The spectra of the ICM in 7 regions were extracted and then fitted together (excluding \# 6 as it covers the curved part of the X-ray tail) in order to increase the signal to noise ratio, which gives a temperature of 7.08$^{+0.32}_{-0.29}$ keV and an abundance of 0.13$\pm 0.06$ $Z_{\odot}$. With the abundances fixed at this overall value, the spectra from the 7 individual regions were fitted separately. The results are shown in the lower right panel of Figure 4. Clearly, region \# 6 has the lowest temperature, which is probably due to the contamination by the remaining cold ISM.  This suggests that the tail is long and continues out to this region.

In order to derive the properties of the ICM near \gab , we estimated the density of the ICM using the $\beta$ model profile for the cluster, $n_e=n_{e,0}(1+r^2/r_c^2)^{-3\beta/2}$, derived by B$\ddot{\rm o}$hringer et al.\ (1996). The cluster center was taken to be the  X-ray surface brightness peak (16:14:22, -60:52:20) as given in the same paper. The parameters for A 3627 are $n_{e,0}=2.4\times10^{-3}$ cm$^{-3}$, $r_c=9'.95$, and $\beta=0.555$. Assuming the cluster-centric distance of \gab\ is the projected distance (110 kpc, see Section 5.1 for a further discussion), the electron density of the ICM at the position of the galaxy is $1.9\times10^{-3}$ cm$^{-3}$. The thermal pressure and ram pressure acting on \gab\ are $4.2\times10^{-11}$ dyn cm$^{-2}$ and $15.3({v_{\rm gal}}/{2000 \rm \ km\ s^{-1}})^2\times10^{-11}$ dyn cm$^{-2}$, respectively.

\section{Discussion}

\subsection{Properties of \gab 's Tails}

The tail of \gab\ is single-sided and our deep SOAR $i$-band image does not reveal any tidal features, so tidal interactions can be ruled out. Ram  pressure stripping is the most likely mechanism to produce the observed tail. A3627 is not relaxed and is probably undergoing a merger with a subcluster to the southeast of the main body of the cluster (B$\ddot{\rm o}$hringer et al.\ 1996; Nishino et al.\ 2012). Thus, the dynamics of the galaxies and the ICM in the cluster are complex, which is reflected by the projected configuration of \gab 's tail with respect to the cluster and the galaxy. The angle between the X-ray tail and the direction to the cluster center (taken as the X-ray peak) is $\sim60^\circ$, so the galaxy is currently moving away from the cluster center in the plane of the sky. Meanwhile, the line-of-sight velocity of the galaxy with respect to the cluster is 872 km s$^{-1}$, suggesting that along the line-of-sight, the galaxy is either approaching the cluster center from the front, or moving away from the cluster center from the behind. The orbit of the galaxy can be constrained further by the early stage of stripping of \gab\  as discussed below. Targeted simulations will be required to fully explore the orbit of \gab , but such simulations are beyond the scope of this paper. 

Despite the complexity of the dynamics, two observational results indicate that most of \gab 's current motion is probably in the plane of the sky. First, the 3-dimensional velocity dispersion in the inner $1/3$ Abell radius (0.67 Mpc) of the cluster is $\sim1700$ km s$^{-1}$ (Woudt et al.\ 2008).  Taking this value as the 3-dimensional velocity of the galaxy, the velocity in the plane of the sky is $1460$ km s$^{-1}$, significantly larger than the line-of-sight velocity ($872$ km s$^{-1}$). Second, the leading edge is very sharp (Figures 1 and 2), suggesting that we are viewing the stagnation region directly.  If so, the angle between the motion and the plane of the sky cannot be large (e.g., Mazzotta et al.\ 2001; Markevitch \& Vikhlinin 2007). As the direction of the tail is opposite to the motion, the tail probably lies approximately in the plane of the sky, so the length of the tail should be close to its projected value. It is also likely that \gab\ is closer to the center of the cluster than \ga, and is, in fact, not too far beyond its projected radius. A comparison of the surface brightness of the X-ray tails of \gab\ and \ga\ (Figure 2 and Figure 5 of S10) shows that the emissivity of the gas in \gab 's tail is higher than that in \ga 's tail. This implies that the pressure around \gab 's tail is higher than that around \ga 's tail (Tonnesen \& Bryan 2011). Thus, \gab\ is probably closer to the cluster center than \ga. Since the cluster-centric distance of \ga\ is close to 180 kpc (S10), this means that the cluster-centric distance of \gab\ is probably less than 180 kpc (the projected distance is 110 kpc). Therefore, both the length of the tail and the cluster-centric distance of \gab\ should be close to their projected values.

The morphology of the X-ray tail (Figure 1) is affected by the galactic potential, the strength of the ram pressure and the cluster gravitational field. The swollen part of the tail may largely be due to the complex potential of the galaxy, while the second straight part should be aligned with the direction of the ICM flow relative to the galaxy (the wind). The first straight part is a consequence of the wind's ram pressure and the restoring force of the galaxy. To push gas outward in the disk, the drag force per unit mass needs to exceed the central acceleration $v_K^2 / r$, where $v_K$ is the circular velocity and $r$ is the distance to the center of the galaxy. The restoring force per unit mass perpendicular to the disk is $2 \pi G \Sigma$, where $\Sigma$ is the surface mass density of the disk. Therefore, if the drag force per unit mass $F$, acts at an angle $\theta$ to the plane of the disk, at the point where the gas just leaves the disk, we have $F \cos\theta \simeq v_K^2 / r$ and $F \sin\theta \simeq 2 \pi G \Sigma$.  Eliminating $F$, this gives the surface density of the disk at that point, $\Sigma \simeq v_K^2 \tan\theta / (2 \pi G r)$.  Taking $\theta \simeq 11^\circ$ and $r \simeq 5$ kpc from Figure 1, and assuming $v_K = 200$ km s$^{-1}$, we get $\Sigma \simeq 57$ $M_{\rm \odot}$ pc$^{-2}$, which is close to the surface mass density ($49\pm6$ $M_{\rm \odot}$ pc$^{-2}$) that Binney \& Tremaine (2008) give for the Milky Way in the solar neighborhood. This agreement is expected, as \gab\ is a Milky-Way-like galaxy and the distance of the point where the gas just leaves the disk to the galaxy's center is close to the Sun's distance to the Galactic center, so the resulting stellar surface density should be comparable. This in turn implies that our simple argument is reasonable. The curved part is probably caused by sinking of the stripped gas due to the gravitational field of the cluster.

The mass of the X-ray tail can be estimated if some approximations are made. If we assume the configuration of the tail can be represented by a cylinder with a length of 40 kpc and radius of 1.5 kpc, and the gas in the tail is homogeneous and the emission can be well described by the single-$kT$ model, utilizing the normalization of the spectral fit, we can derive a total mass of $2.0f^{1/2}\times10^8$ $M_{\rm \odot}$, where $f$ is the filling factor of the soft X-ray emitting gas. Applying the same approach to the individual regions gives a total mass of $2.5f^{1/2}\times10^8$ $M_{\rm \odot}$. This is a lower limit of the total mass of the tail, as we only account for the X-ray bright part of the tail and this mass is only the mass of the X-ray emitting gas. S10 estimated a total ISM mass of the galaxy to be $\sim(4-9)\times10^9$ $M_{\rm \odot}$ initially. Thus, the X-ray tail accounts for $2.2f^{1/2}\%-5.0f^{1/2}\%$ of the initial ISM mass, which is very small given the high ram pressure. Numerical simulations often show that large fractions (80\% or even higher) of the gas are stripped out of galaxies (e.g., Kapferer et al.\ 2009; Tonnesen \& Bryan 2009) after a sufficiently long time. The low mass in the X-ray tail suggests that a large fraction of the ISM may be still in the disk. Actually, with two pointings toward \gab , J\'achym et al.\ (2013, in preparation) found $\sim3.5\times10^9$ $M_{\rm \odot}$ of molecular gas in the galaxy with the Atacama Pathfinder EXperiment telescope (APEX). Thus, \gab\ is probably at an early stage of stripping. 

The mass of the H$\alpha$ tails can be estimated as follows. The H$\alpha$ tails can be approximated as three cylinders centered at (16:13:36.53, -60:52:13.4), (16:13:38.06, -60:52:43.7), and (16:13:39.42, -60:52:41.9) (J2000), respectively. The dimensions of the three cylinders are $1.5\times7.6$ kpc, $1\times13.5$ kpc, and $1\times12$ kpc (radius$\times$length), respectively. While the first cylinder is still in the stellar disk of the galaxy, the other two cylinders represent the two tails after the bifurcation point and are beyond the stellar disk. The surface brightness of the H$\alpha$ emission was measured. We corrected the H$\alpha$ flux for the [N II] lines by assuming that the combined strength of the [N II] lines was 30\% of the H$\alpha$ line. No intrinsic absorption was assumed. The average H$\alpha$ surface brightness in the three regions is 10$\times10^{-17}$ erg s$^{-1}$ cm$^{-2}$ arcsec$^{-2}$, 2.3$\times10^{-17}$ erg s$^{-1}$ cm$^{-2}$ arcsec$^{-2}$, and 1.0$\times10^{-17}$ erg s$^{-1}$ cm$^{-2}$ arcsec$^{-2}$, respectively. The total H$\alpha$ flux from the three regions is 2.9$\times10^{-14}$ erg s$^{-1}$ cm$^{-2}$, but 73\% of the flux comes from the first region which is still in the stellar disk. This total flux can be compared with that from \ga 's H$\alpha$ tails, 4.4$\times10^{-14}$ erg s$^{-1}$ cm$^{-2}$ (S07). To estimate the mass of the H$\alpha$ emitting gas, we assume an electron temperature of $10^4$ K and an effective recombination coefficient for H$\alpha$ of $1.17\times10^{-13}$ cm$^3$ s$^{-1}$ from the case B nebular theory (Osterbrock \& Ferland 2006). The derived electron density is $0.05f_{\rm H\alpha}^{-1/2}-0.13f_{\rm H\alpha}^{-1/2}$ cm$^{-3}$, where $f_{\rm H\alpha}$ is the volume filling factor of the H$\alpha$ emitting gas. The total mass is $4.2f_{\rm H\alpha}^{1/2}\times10^8$ $M_{\rm \odot}$ (the mass of the three parts is $2.4f_{\rm H\alpha}^{1/2}\times10^8$ $M_{\rm \odot}$, $1.1f_{\rm H\alpha}^{1/2}\times10^8$ $M_{\rm \odot}$, and $0.67f_{\rm H\alpha}^{1/2}\times10^8$ $M_{\rm \odot}$, respectively). Numerical simulations of ram pressure stripping indicate that bright H$\alpha$ emission is produced at the edges of dense neutral clouds (Tonnesen \& Bryan 2011). Thus, the volume filling factor should be small. While the observed root mean square (rms) electron density in the tail of \ga\ is $\sim0.045f'^{-1/2}_{\rm H\alpha}$ cm$^{-3}$ (S07), in simulations (with the ICM thermal pressure consistent with that around \ga ), the typical local density of the H$\alpha$ gas is $\sim1$ cm$^{-3}$ (Tonnesen \& Bryan 2011). Thus, the volume filling factor in this case is about 0.2\%. If this filling factor is adopted, the total mass of the H$\alpha$ emitting gas in \gab 's tails is $1.9\times10^7$ $M_{\rm \odot}$. Of course, the volume filling factor is a big uncertainty here. Nevertheless, this suggests that the mass of the H$\alpha$ tails is probably a small fraction of the X-ray tail. H I observations failed to detect H I emission toward the direction of \gab\ (Vollmer et al.\ 2001b), but \gab\ is only 8$'$ from the strong radio galaxy PKS~1610–60 (43 Jy at 1.4 GHz). Future deep multiwavelength observations are required to constrain the amount and distribution of the multiphase gas in the tail.

\subsection{Is the X-ray Tail Over-Pressured?}

The thermal pressure in the X-ray tail can be calculated by spectral fitting. From the single-$kT$ model fit, the electron density in the tail is $0.025f^{-1/2}$ cm$^{-3}$. For a temperature of 0.95 keV, this yields a thermal pressure of $7.3f^{-1/2}\times10^{-11}$ dyn cm$^{-2}$ in the tail. Pressures in individual regions range from $4.6f^{-1/2}\times10^{-11}$ dyn cm$^{-2}$ to $9.4f^{-1/2}\times10^{-11}$ dyn cm$^{-2}$, with $\sim30\%$ errors. We also fitted a double-$kT$ model. By assuming pressure balance between the higher and lower temperature components in the tail, we find that the higher temperature component occupies 94\% of the total volume of the stripped gas. Taking into account the normalizations of the two components, we obtain densities of $0.020f^{-1/2}$ cm$^{-3}$ and $0.054f^{-1/2}$ cm$^{-3}$ for the hotter and cooler components, respectively (note that $f$ is the total volume filling factor of the two components, not the factor for each component separately). This yields a thermal pressure of $6.6f^{-1/2}\times10^{-11}$ dyn cm$^{-2}$ in the tail. Finally, we tried a multi-$kT$ model, which gives a total pressure of $17.8f^{-1/2}\times10^{-11}$ dyn cm$^{-2}$ under isobaric condition. We list the thermal pressures in the tail with different models together with the pressure in the ICM and the ram pressure in Table 3. We can see that the tail is ``over-pressured'' for all three models. 

The nearly constant width along the length ($\gtrsim 40$ kpc) of the tail implies that it should be in near pressure equilibrium with the surrounding ICM, rather than over-pressured. Otherwise, the tail would expand with distance. There are several effects that might explain, at least partially, the apparent higher pressure in the tail \footnote{ Note that simply abandoning the assumption that the cluster-centric distance is close to the projected distance does not ease this problem, since it results in a larger cluster-centric distance and reduces the pressure in the ICM, which makes the ``over-pressure problem'' more severe.}. 

First, the normalization of the spectral fit, and hence the electron density and pressure, is very sensitive to the unknown abundance. This is due to the fact that at these temperatures, most of the X-ray emission is due to lines from heavy elements. For example, with the single-$kT$ fit, when the abundance is 0.18 $Z_{\rm \odot}$ (best-fit), 0.5$Z_{\rm \odot}$, and 1$Z_{\rm \odot}$, respectively, the pressure in the tail is $7.3f^{-1/2}\times10^{-11}$ dyn cm$^{-2}$, $5.2f^{-1/2}\times10^{-11}$ dyn cm$^{-2}$, and $3.8f^{-1/2}\times10^{-11}$ dyn cm$^{-2}$, accordingly. Note that a higher abundance, which is more reasonable for \gab , significantly reduces the gas pressure in the tail. 

Second, the lower thermal pressure in the ICM may imply that other sources of pressure are important (e.g., magnetic pressure). Adopting $7.3f^{-1/2}\times10^{-11}$ dyn cm$^{-2}$ and $4.2\times10^{-11}$ dyn cm$^{-2}$ for the pressures in the tail and in the ICM, respectively, and assuming a filling factor of 1, we acquire a magnetic field strength of $\sim30$ $\mu$G at the position of the tail in A3627, which is larger than typical values of a few $\mu$G in clusters.  It is representative of the magnetic fields in the very centers of cool-core clusters (e.g., Carilli \& Taylor 2002), but A3627 is a non-cool-core cluster (B$\ddot{\rm o}$hringer et al.\ 1996). We emphasize that it is unlikely that the magnetic field in the ICM is this large at this location. Moreover, we did not account for magnetic support in the tail, which may be amplified in the process of stripping and might even dominate (Ruszkowski et al.\ 2012). Therefore, the true magnetic field strength may be very different from the above value. Nevertheless, magnetic pressure in the ICM might help to alleviate the ``over-pressure problem".

Last but not least, so far we have assumed that the soft X-ray emission is of thermal origin, but charge exchange may also produce a fraction of the observed soft X-ray flux \footnote{ Charge exchange may also produce H$\alpha$ emission, but we only introduce it here as a possible mechanism to produce some of the soft X-ray emission, in order to explain the apparent higher pressure in the tail. A thorough study and detailed modeling of charge exchange in the case of stripped tails is beyond the scope of this paper.} (see Dennerl 2010 and Lallement 2004 for a review and application to astrophysical environments). Due to the poor spectral quality of the data and many unknown quantities, it is impossible to resolve individual lines and calculate the contribution to the total flux from charge exchange emissions. Nevertheless, a crude upper limit can still be obtained if some assumptions and approximations are made. The local emissivity of charge exchange for a particular line (assuming single electron capture) is $P\propto n_i n_{\rm neu}V\sigma$, where the four quantities from left to right are the ion number density, the neutral gas density, average relative velocity of the ions and neutral atoms, and the cross section of charge exchange emission for this particular line (Liu et al.\ 2011). H I observations with a column density sensitivity limit of $2\times10^{20}$ cm$^{-2}$ failed to detect H I toward \gab\ (Vollmer et al.\ 2001b), from which we derive an upper limit on the average H I gas density in the tail of $\lesssim 0.02$ cm$^{-3}$. As the X-ray emission from the tail is most significant in the $0.7-1.2$ keV band, following Fabian et al.\ (2011), we only account for charge exchange emissions from Fe L-shell lines and O VIII lines in the same energy band and assume that they all have an average energy of 1 keV. The relative velocity is another unknown parameter. We assume that once the neutral gas has been stripped, it is at rest relative to the ICM and the velocities of the ions are their rms thermal velocities. In this way, we obtain an upper limit on the  flux from charge exchange emissions of $\lesssim 3\times10^{-10}$ erg s$^{-1}$ cm$^{-2}$. This limit is almost four orders of magnitude larger than the observed flux in the $0.7-1.2$ keV band. Note that in the above crude upper limit, we used the whole tail volume, which is much larger than the true volume where charge exchange takes place, as this process only occurs roughly at the interface of the hot and cold gas. So the true flux due to charge exchange would be much smaller than this limit.  Nevertheless, this suggests that charge exchange may potentially contribute to the observed soft X-ray flux, and hence reduce the derived pressure in the tail.

In summary, the uncertainties in the abundance, thermal vs. non-thermal X-ray emission, or magnetic support in the ICM may cause the pressure in the tail to appear to be higher than that in the ICM. However, the nearly constant width of the tail suggests that it is most likely in near pressure equilibrium with the ambient ICM.

\subsection{Comparison with Other Observations of Galaxy Tails}

Despite growing observational evidence for tailed galaxies in different bands, X-ray tails are rare for late-type galaxies, and  their formation mechanisms appear to be quite diverse. There is a feature extending about 22$''$ (88 kpc) attached to C 153 in A2125 (Wang et al.\ 2004). Because of the limited statistics, it is not clear whether the feature indeed represents a coherent diffuse X-ray tail. UGC 6697, a starburst galaxy, has an X-ray tail up to 60 kpc in length, but most of its X-ray emission is within the galaxy (Sun \& Vikhlinin 2005). Its velocity field and peculiar morphology may imply the presence of a second galaxy hidden behind the main body of UGC 6697 (Gavazzi et al.\ 2001b). Thus, tidal interaction may also contribute to its formation. NGC 6872 is the first galaxy with a long (90 kpc) X-ray tail in a poor galaxy group. The tail lies in the region between NGC 6872 and the dominant elliptical galaxy NGC 6876 in the Pavo group.  This tail might be either the intragroup gas gravitationally focused into a Bondi-Hoyle wake, or a thermal mixture of the intragroup gas with gas removed from NGC 6872 by turbulent viscous stripping, or a combination of both (Machacek et al.\ 2005). NGC 4388, NGC 4501, NGC 4438 and NGC 2276 all possess short ($10-20$ kpc) X-ray tails (Machacek et al.\ 2004; We$\dot{\rm z}$gowiec et al.\ 2011; Rasmussen et al.\ 2006), while most of the diffuse X-ray emission from NGC 4388 and NGC 4438 splits into two main streams that may represent double tails (Randall et al.\ 2008). Both \ga\ and \gab\ have long X-ray tails ($\sim80$ kpc and $\gtrsim40$ kpc, respectively) and double tails (X-rays and H$\alpha$ for \ga , H$\alpha$ as well as an X-ray enhancement apart from the main X-ray tail for \gab ). Moreover, both galaxies reside in the same cluster. We mainly compare the tails of these two galaxies below. S07 and S10 studied the tails of \ga\ in detail. For simplicity, we will not list the references when we cite results from them except for special emphasis. The reader can find more information about \ga\ and its tails in these two papers. The X-ray properties of \ga\ and \gab\ as well as their tails are summarized in Table 4, in which we also update the results for \ga 's X-ray tails with AtomDB v2.0.1 in order to allow direct comparisons.
 
\ga\ is a blue emission-line galaxy at a projected distance of 180 kpc from the X-ray peak of the cluster. Its radial velocity is close to the average velocity of the cluster (Woudt et al.\ 2008), which means that most of its motion is probably in the plane of the sky. \gab\ is redder and more massive with a projected distance of 110 kpc to the cluster center. As discussed in Section 5.1, most of its motion is also probably in the plane of the sky. There are no signs of merger or tidal features for either galaxy. Therefore, the tails of both galaxies are formed due to ram pressure stripping. From the directions of the tails and morphologies of the galaxies, it is inferred that \ga\ is probably undergoing nearly face-on stripping while \gab\ is undergoing nearly edge-on stripping. 

The length of \ga 's main X-ray tail is twice as long as the significantly detected part of \gab 's X-ray tail. Both tails are very narrow, although \gab 's tail is twice as thin, which may be due to the nearly edge-on stripping for \gab\ compared to the nearly face-on stripping for \ga . The width do not increase along the tail for either galaxy, indicating that the tail should be in near pressure equilibrium with the surrounding ICM. However, X-ray spectral fitting yields apparently higher pressures than in the surrounding ICM (see Section 5.2 for a detailed discussion). There are H$\alpha$ tails for both galaxies, while \ga\ also has a H$_2$ tail (Sivanandam et al.\ 2010).  However, these H$\alpha$ and H$_2$ tails have masses which are probably much smaller than the X-ray tails. The low mass of the X-ray tail compared to the initial mass of the ISM for \gab\ may indicate that this galaxy is at an early stage of stripping, which is confirmed by a large amount of molecular gas still remaining in its disk (J\'achym et al.\ 2013, in preparation).

Both X-ray tails of \ga\ and the X-ray tail of \gab\ have nearly constant temperatures along their lengths (0.92 keV and 0.95 keV, respectively), and the temperatures of the two galaxies' tails are very similar ($\sim1$ keV) which may suggest that some properties of the tail might be somehow affected by the cluster environment. Moreover, the abundances with the single-$kT$ fits of the tails of the two galaxies are very low, likely a reflection of multiphase gas in the tails.

A striking difference between the H$\alpha$ tails of the two galaxies is that \ga\ has over 30 luminous H II regions in its stripped gas while none was detected in \gab 's tails (S07; S10). To search for H II regions downstream \gab 's H$\alpha$ tails, we followed a similar analysis as S07 for \ga . SExtractor was run to detect compact sources like H II regions and stars in the two narrow bands (S07), giving the noise level of the data. The same color selection criteria as used in S07 (Section 3) were applied, which results in zero detection. The upper limit on the H$\alpha$ luminosity for an individual H II region that was undetected is 10$^{38}$ erg s$^{-1}$. In comparison, there are 27 H II regions above this limit detected in \ga 's H$\alpha$ tails, with a total SFR of 0.25 $M_{\rm \odot}$ yr$^{-1}$ (no intrinsic absorption assumed). It is unclear how many undetected H II regions there are in \gab 's H$\alpha$ tails. However, the average H$\alpha$ surface brightness of the parts of \gab 's tails that are beyond the galactic disk is $\sim 1.5\times10^{-17}$ erg s$^{-1}$ cm$^{-2}$ arcsec$^{-2}$, similar to the typical H$\alpha$ surface brightness of \ga 's tails and other H$\alpha$ tails (S07; Gavazzi et al.\ 2001a; Yagi et al.\ 2007). The total H$\alpha$ luminosity of the parts of the tails beyond the galactic disk is 4.5$\times10^{39}$ erg s$^{-1}$. The excitation mechanism of the H$\alpha$ tails is complex and beyond the scope of this paper. Even if all the diffuse H$\alpha$ emission from the tails beyond the galactic disk is caused by star formation, assuming the Kennicutt (1998) SFR$-L_{\rm H\alpha}$ relation and no intrinsic extinction, the total SFR is only 0.036 $M_{\rm \odot}$ yr$^{-1}$. Therefore, the star formation activity in \gab 's tails is much weaker than that in \ga's tails. If the cluster-centric distances of both galaxies are close to their projected distances, as discussed above and in Section 5.1, then both the density and pressure around \gab 's tails are higher than those around \ga 's tails (Table 4), which should advantage star formation in \gab 's tails (Kapferer et al.\ 2009; Tonnesen \& Bryan 2012). What mechanism prohibits it then? Is the surface density of cold gas in \gab 's tails too low to form stars? What is the critical factor that determines SFR in a tail? These questions cannot be answered with current data.

The most striking similarity between the tails of \ga\ and \gab\ is that they both show double-tail features (see S10 for the case of \ga ). The double tails of \ga\ are seen in X-rays and H$\alpha$, while the double tails of \gab\ are in H$\alpha$ (although they are probably also double in X-rays). One apparent difference is that the secondary tails bifurcate at different distances from the nuclei of the two galaxies (see Figure 1 in this work and Figure 1 in S10). The formation mechanism of the double tails is not known. Two possible explanations are discussed in the next section.

\subsection{Comparison with Numerical Simulations}

Since the first analytical discussion by Gunn \& Gott (1972), many theoretical studies of ram pressure stripping have been done with more realistic conditions, such as the inclusion of varying winds, viscosity, multiphase ISM, radiative cooling, and star formation (Abadi et al.\ 1999; Schulz \& Struck 2001; Vollmer et al.\ 2001a; Marcolini et al.\ 2003; J\'achym et al.\ 2007; Roediger \& Br$\ddot{\rm u}$ggen 2008a; Kapferer et al.\ 2009; Tonnesen \& Bryan 2009, and references therein). In general, these simulations produce truncated gas disks down to the stripping radius and long tails in the downwind directions which are qualitatively consistent with observations, but a complete quantitative agreement among the simulations or with observations has not been achieved yet.

Tails in the simulations are generally much longer (e.g., $>$ 100 kpc in Roediger \& Br$\ddot{\rm u}$ggen 2008b and Tonnesen \& Bryan 2010, 400 kpc in Kapferer et al.\ 2009), more flared, and wider (e.g., 85 kpc at a distance of 125 kpc to the galaxy in Tonnesen \& Bryan 2010). The simulations by Roediger \& Br$\ddot{\rm u}$ggen (2006) show that the widening of the tails is independent of the galactic inclination, which led them to conclude that the dynamics of the widening is independent of the galaxy but intrinsic to the ICM flow. Turbulence in the wake is the main cause of the flaring (Roediger \& Br$\ddot{\rm u}$ggen 2008b). Given the nearly constant width of \gab 's X-ray tail, turbulence must not be very strong in the tail. Including radiative cooling results in tails that are significantly narrower and less flared (Tonnesen \& Bryan 2010), which are in better agreement with observations. To compare with \ga , Tonnesen \& Bryan (2011) used ICM parameters comparable to those around this galaxy. Tails with nearly constant widths could be produced. However, the tails were too wide in their simulations. They did model a larger galaxy than \ga ;  however, the widths of the tails were still greater than the diameter of the galaxy, which is inconsistent with observations of \ga\ and its tails. Because they modeled a face-on stripping, we cannot compare their simulations directly with the X-ray tail of \gab .  However, the extreme narrowness of \gab 's tail is likely to be a big challenge for simulations, considering the difficulty they had to produce a narrow tail comparable to observations for face-on stripping. 

What is the relation between tails in different bands (e.g., X-rays, H$\alpha$, H I)? What is the distribution of the multiphase gas? What determines the survival of the stripped gas? Tonnesen \& Bryan (2011) simulated ICM conditions similar to those around \ga\ to investigate these questions. Their results show that a high ICM pressure ($>9\times10^{-12}$ erg cm$^{-3}$ for the conditions in their simulations) will result in an X-ray bright tail. From Table 4, both the ICM pressures around \ga\ and \gab\ are higher than the minimum pressure that is required to make the tail X-ray bright in their simulations. The X-ray emission comes from mixing of cold gas stripped from the galaxy with the hot ICM and is not localized near the dense clouds, in contrast to the H$\alpha$ emission which peaks at the edges of the neutral clouds so it traces their distribution (Tonnesen \& Bryan 2011). This distribution of the multiphase gas is consistent with the longer X-ray tails than H$\alpha$ tails observed for both \ga\ and \gab\ (Tonnesen \& Bryan 2011). However, they only modeled mixing via adiabatic compression and small-scale turbulence. Thermal conduction, which they did not include, may transport a large amount of heat from the ICM to the cold stripped gas; this would have a profound impact on the phase diagram of the gas in the tails. H I tails were predicted for both \ga\ and \gab , with at least the same lengths as their H$\alpha$ tails. Therefore, future deep 21-cm observations and observations in other bands (e.g., CO observations) would test their model and better constrain the distribution of the multiphase gas in the tails, as well as the physical processes that determine the excitation and survival of the tails in different bands.

Although ram pressure acts to strip gas down to the stripping radius, which will eventually quench star formation, some authors argue that the central region of the galaxy will be compressed so as to increase star formation or even induce a star burst on a shorter timescale (Fujita \& Nagashima 1999; Schulz \& Struck 2001; Vollmer et al.\ 2001a; Kronberger et al.\ 2008; Kapferer et al.\ 2009; but see Tonnesen \& Bryan 2012). Star formation in the stripped tail can also occur as revealed both observationally (e.g., S07; Yoshida et al.\ 2008) and theoretically (e.g., Kapferer et al.\ 2009; Tonnesen \& Bryan 2012). Kapferer et al.\ (2009) applied a combined $N$-body/hydrodynamic description (GADGET-2) with radiative cooling and a recipe for star formation and stellar feedback to calculate the effect of ram pressure stripping on disk galaxies. Their results suggest that star formation could be enhanced by more than an order of magnitude under a high ram pressure (5$\times10^{-11}$ dyn cm$^{-2}$), and that up to 95\% of the newly formed stars can be found in the wake of the galaxy out to a distance greater than 350 kpc. The enhancement in star formation is more dependent on the surrounding gas density than on the relative velocity (Kapferer et al.\ 2009). On the other hand, Tonnesen \& Bryan (2012) used the adaptive mesh refinement (AMR) code ENZO including cooling and thermal feedback to study star formation in ram pressure stripped tails and found that SFR in the tail is low and depends primarily on the pressure in the ICM rather than the ram pressure strength. 

Despite the many differences between their methods and results, both groups agree that a higher ICM density or pressure will increase star formation in the tail. From Table 4, we can see that both the ICM density and pressure around \gab 's tails are higher than those around \ga 's tails (assuming that the cluster-centric distances of the two galaxies are close to their projected distances, see the discussions in Sections 5.1 and 5.3), yet star formation is far more active in \ga 's tails than in \gab 's tails (Section 5.3). This inconsistency with those simulations may imply that, due to the complexity of star formation, the effects of some other factors (e.g., thermal conduction or magnetic fields) exceed those of external pressure and density on star formation in the tails of the two galaxies in question. 

One particular factor we would like to address is inclination. \gab\ is undergoing nearly edge-on stripping, in which case continuous stripping (Nulsen 1982) is important and stripping proceeds on a longer time-scale (see Roediger \& Br$\ddot{\rm u}$ggen 2006 and references therein). It is reasonable to assume that the gas is more uniformly deposited in the wake due to the much milder stripping. This is supported by the smooth morphology of \gab 's X-ray tail compared to the clumpy X-ray tails of \ga\ which is undergoing nearly face-on stripping. One might expect that star formation is more difficult in a tail with a more uniform gas distribution, which may (partly) explain the difference between the tails of \ga\ and \gab . The test of this hypothesis and the answer to the issue will require further study.

The origin of the double-tail feature is an interesting question. S10 presented the discovery of the first long double X-ray tails associated with a late-type galaxy, \ga , and proposed that stripping of two spiral arms results in the double tails. While this simple explanation is appealing because gas indeed follows spiral structures in late-type galaxies, it may have a problem in explaining the double tails in face-on stripping cases. None of the purely hydrodynamical simulations of ram pressure stripping have produced double tails. Recently, Ruszkowski et al.\ (2012) made the first simulations of a disk galaxy exposed face-on to a uniformly magnetized wind. The inclusion of magnetic fields has a strong impact on the morphology of the tails; they are more filamentary rather than clumpy in the hydrodynamical simulations. Their simulations also show the formation of double magnetized density tails which Ruszkowski et al.\ (2012) interpreted as the folding of the ambient magnetic fields around the galaxy. However, more work needs to be done to compare this folding of the magnetic field with previous studies on magnetic draping (e.g., Lyutikov 2006; Dursi \& Pfrommer 2008; Pfrommer \& Dursi 2010). Future simulations should consider a varying wind, as well as include the effect of the interstellar (as opposed to intracluster) magnetic field, which evolves due to compression and shear motions (e.g., Otmianowska-Mazur \& Vollmer 2003) or dynamo processes (e.g., Moss et al.\ 2012) induced by the ram pressure. They should also consider the back reaction exerted by the magnetic field on the process of stripping (e.g., magnetic draping). Nevertheless, the results of Ruszkowski et al.\ (2012) are intriguing and suggest that magnetic field are important in ram pressure stripping. More targeted simulations including both the intracluster and interstellar magnetic fields as well as modeling of spiral arms are needed for a better understanding of the formation of double tails.

\section{Summary}

In this work, we present the analysis of a new, on-axis observation of \gab\ with the \chandra\ X-ray observatory, as well as the results of optical observations with SOAR. We study the spatial structure and spectral properties of the tails. Detailed comparisons with the similar tails of \ga\ and with simulations are made. Our main results are:
  
1) The \chandra\ data show a long ($\gtrsim$ 40 kpc) and narrow tail with a nearly constant width ($\sim3$ kpc) downstream of \gab\ (Section 3.1 and Figures 1 and 2) . The H$\alpha$ image also reveals a tail spatially coincident with the X-ray tail, with a sharp leading edge corresponding to the X-ray edge (Section 3.1 and Figures 1 and 3). We conclude that the tail is caused by nearly edge-on stripping by ram pressure (Sections 5.1 and 5.3).

2) The X-ray tail has a temperature of 0.95$^{+0.04}_{-0.05}$ keV with the single-$kT$ fit (Section 3.2 and Table 2). The abundance with the single-$kT$ fit is low (0.18$^{+0.09}_{-0.05}$ $Z_{\odot}$), which indicates that the gas in the tail is multiphase (Section 3.2). Spectral fits to individual regions along the tail with the single-$kT$ model give similar temperatures (Figure 4). The spectral fit to the nucleus reveals an obscured Seyfert-2-like AGN with a bolometric luminosity of $\sim1.2\times10^{43}$ erg s$^{-1}$ (Section 3.2).

3) The mass of the X-ray tail is only a small fraction ($<5\%$) of the initial ISM mass of the galaxy (Section 5.1). This suggests that the stripping of \gab 's ISM may be at an early stage, which is confirmed by the detection of a large amount of molecular gas in the disk. The mass of the H$\alpha$ tails is probably a small fraction of the X-ray tail (Section 5.1). Future H I, CO, and infrared observations of \gab\ are required to constrain the  amount and distribution of the ISM both in the galaxy and its tail.
 
4) X-ray spectral fitting with the single-$kT$, double-$kT$, and multi-$kT$ models yields apparently higher pressures in the tail than in the nearby ICM (Section 5.2 and Table 3). However, the nearly constant width along the tail suggests that it is in near pressure equilibrium with the ambient ICM. This ``over-pressure problem" may be due to the sensitive dependence of pressure on the abundance, or additional source of pressure (e.g., magnetic field) in the ICM, or non-thermal origin (i.e., charge exchange) of part of the observed soft X-ray emission (Section 5.2). Deep X-ray spectroscopic data and radio observations are needed to place better constraints. The nearly constant width also disfavors strong turbulence in the tail (Section 5.4). 

5) Apart from an H$\alpha$ tail ($\sim21$ kpc) corresponding to the X-ray tail, the SOAR data also reveal a secondary tail ($\sim12$ kpc) bifurcating at a distance of $\sim7.5$ kpc from the nucleus at a $\sim23^\circ$ angle to the main tail (Section 3.1 and Figure 1). Along the secondary H$\alpha$ tail, the X-ray surface brightness is enhanced at the $\sim2 \sigma$ level. Future deep X-ray observations are required to examine the possible double X-ray tails of \gab .

6) Comparisons between the tails of \gab\ and \ga\ show similarities (e.g., narrow, not flared, constant temperature, double tails) and differences (e.g., star formation, tail morphology) (Section 5.3), both of which pose challenges to current simulations (Section 5.4).  Why are the tails so narrow?  What effect does thermal conduction have on the multiphase gas in the tails? Why are X-ray tails so rare for late-type galaxies? What are the crucial factors that determine SFR in the tails, and why do these differ between \ga\ and \gab ? What is the origin of the observed double tails?

The discovery of \gab 's tails (X-rays and H$\alpha$) provides a unique laboratory to study ram pressure stripping and the subsequent gas evolution in the tails. First, \gab\ is undergoing nearly edge-on stripping, while most simulations deal with a face-on geometry. Consequently, it is a good case to study the influence of inclination on the behavior of ram pressure stripping, including mass loss (Roediger \& Br$\ddot{\rm u}$ggen 2006), the gas distribution, and SFR in the tail (Section 5.4). Second, it is known that bright X-ray tails are rare for late-type galaxies (S10). The discovery of \gab 's X-ray tail provides new constraints on the origin of X-ray bright tails (Section 5.4), and may provide some clues on why X-ray tails are so rare. Third, the fact that the X-ray tails of both \ga\ and \gab\ have very narrow and nearly uniform widths, constant and similar temperatures, ``low abundances'', and are apparently ``over-pressured'' compared to the surrounding ICM (Section 5.3) may indicate that they are common features for X-ray tails of late-type galaxies. The explanation of these features may depend on the details of microphysical processes, which also are the central problems in other astrophysical fields (e.g., cool-cores, AGN feedback, and ICM physics). Fourth, the double-tail features require refined modeling of galaxies (e.g., including spiral arms) and the inclusion of magnetic fields both in the intracluster and interstellar space (Section 5.4). Ultimately, more data and more sophisticated simulations are required to study in detail the stripping process and gas evolution in the stripped tails, which have important implications for galaxy evolution and baryonic physics.

\acknowledgments

We thank Pavel J\'achym for providing the unpublished results of the observations with APEX. We thank Q. D. Wang for stimulating discussions. We thank the referee for thorough and constructive comments, which was very helpful in improving the quality of this paper. The scientific results reported in this article are based on observations made by the \chandra\ X-ray Observatory, ObsIDs 12950 and 9518. We also present data obtained with the Southern Observatory for Astrophysical Research (SOAR) telescope, which is a joint project of the Minist\'{e}rio da Ci\^{e}ncia, Tecnologia, e Inova\c{c}\~{a}o (MCTI) da Rep\'{u}blica Federativa do Brasil, the U.S. National Optical Astronomy Observatory (NOAO), the University of North Carolina at Chapel Hill (UNC), and Michigan State University (MSU). We made use of the NASA/IPAC Extragalactic Database (NED) which is operated by the Jet Propulsion Laboratory, California Institute of Technology, under contract with the National Aeronautics and Space Administration. M. S. is supported by the NASA grants GO1-12103A, GO0-11145C and GO0-11008B. L. J. is supported by the 100 Talents program of Chinese Academy of Sciences (CAS) and the Key Laboratory of Dark Matter and Space Astronomy of CAS. X. K. is supported by the National Natural Science Foundation of China (NSFC, No. 11225315), the Specialized Research Fund for the Doctoral Program of Higher Education (SRFDP, No. 20123402110037), and the Chinese Universities Scientific Fund (CUSF).

{\it Facilities:} \facility{CXO (ACIS)}, \facility{SOAR (SOI)}.

\begin{table}
\begin{center}
{\caption{Basic Parameters of \ga\ and \gab }
\vspace{0.5cm}
\begin{tabular}{lcc} \hline \hline 
 Parameter & \ga\ & \gab\ \\ \hline 
 $D$ (kpc)$^a$  & 180 & 110 \\
 $v_{\rm rad}$ (km s$^{-1}$)$^b$  & 4680 (191) & 5743 (872) \\
 $M_{\rm *}$ ($10^9 \ M_{\rm \odot}$)$^c$ &  $5-8$ & $32-39$ \\
 $L_{K_{\rm s}}$ ($10^{10} \ L_{\odot}$)$^{c}$ & 2.63 (1.51) & 12.88 (15.85) \\
 $B-K_{\rm s}$ (mag)$^{c}$ & 2.76 (2.15) & 3.18 (3.40) \\
 \hline \hline
\end{tabular}
\vspace{-0.5cm}
\tablenotetext{a}{Projected distance to the X-ray peak of the cluster.}
\tablenotetext{b}{Radial velocity from Woudt et al.\ (2004, 2008). Values in parentheses are the differences from the cluster's mean velocity.}
\tablenotetext{c}{The stellar mass, $K_{\rm s}$ band luminosity, and $B-K_{\rm s}$ magnitude, adopted from S10. Values in parentheses are from the Two Micro All Sky Survey (2MASS).}
}
\end{center}
\end{table}

\begin{table}[h] 
\scriptsize
\begin{center}
{\caption{Spectral Fits to the X-ray Tail and the Nucleus$^a$}
\vspace{0.5cm}
\begin{tabular}{llcccc} \hline \hline
Region$^{b}$ & Model$^{c}$ & Parameters$^{d,e}$ & C-statistic$^d$ &  Parameters$^{e,f}$ & C-statistic$^f$ \\
& & (v2.0.1) & (d.o.f) & (v1.3.1)  & (d.o.f) \\  \hline 

Tail & APEC & $kT$=0.95$^{+0.04}_{-0.05}$, $Z$=0.18$^{+0.09}_{-0.05}$ & 53.9 (69)& $kT$=0.80$^{+0.03}_{-0.04}$, $Z$=0.14$^{+0.06}_{-0.04}$ & 53.8 (69)\\[2pt]

& APEC & $kT$=0.98$^{+0.03}_{-0.03}$, $Z$=(1.0)  & 66.8 (70) & $kT$=0.80$^{+0.03}_{-0.03}$, $Z$=(1.0) &  70.1 (70) \\[2pt]

& APEC+APEC & $kT_{\rm 1}$=1.06$^{+0.09}_{-0.07}$, $Z_{\rm 1}$=0.28$^{+0.17}_{-0.09}$ & 49.5 (67) & $kT_{\rm 1}$=1.00$^{+0.10}_{-0.08}$, $Z_{\rm 1}$=0.32$^{+0.24}_{-0.13}$ & 49.0 (67) \\[2pt]
&  & $kT_{\rm 2}$=0.40$^{+0.22}_{-0.13}$, $Z_{\rm 2}$=$Z_{\rm 1}$ &  & $kT_{\rm 2}$=0.43$^{+0.19}_{-0.09}$, $Z_{\rm 2}$=$Z_{\rm 1}$ &    \\[2pt]

& APEC+APEC & $kT_{\rm 1}$=1.05$^{+0.09}_{-0.07}$, $Z_{\rm 1}$=0.27$^{+0.18}_{-0.09}$ &  49.9 (67) &  $kT_{\rm 1}$=1.00$^{+0.09}_{-0.09}$, $Z_{\rm 1}$=0.31$^{+0.28}_{-0.13}$ &  49.4 (67)   \\[2pt]
&  &  $kT_{\rm 2}$=0.39$^{+0.21}_{-0.15}$, $Z_{\rm 2}$=(1.0) & & $kT_{\rm 2}$=0.42$^{+0.17}_{-0.08}$, $Z_{\rm 2}$=(1.0) &  \\[2pt]

& CEMEKL  & $\alpha$=1.55$^{+1.07}_{-0.73}$, $kT_{\rm max}$=1.15$^{+0.46}_{-0.15}$, $Z$=0.30$^{+0.63}_{-0.11}$  & 49.6 (68) & $\alpha$=1.53$^{+1.09}_{-0.72}$, $kT_{\rm max}$=1.14$^{+0.46}_{-0.15}$, $Z$=0.30$^{+0.64}_{-0.10}$ & 49.6 (68) \\[2pt]

& CEMEKL  & $\alpha$=1.00$^{+0.45}_{-0.38}$, $kT_{\rm max}$=1.55$^{+0.37}_{-0.24}$, $Z$=(1.0)  & 50.8 (69) & $\alpha$=1.03$^{+0.43}_{-0.40}$, $kT_{\rm max}$=1.52$^{+0.39}_{-0.22}$, $Z$=(1.0)  & 50.8 (69)  \\[2pt]

& CEMEKL  & $\alpha$=0.01$^{+0.05}_{-0}$, $kT_{\rm max}$=(7.0), $Z$=(1.0)  & 66.0 (70) & $\alpha$=0.01$^{+0.05}_{-0}$, $kT_{\rm max}$=(7.0), $Z$=(1.0)  & 66.0 (70)  \\[2pt]

Head & APEC & $kT$=0.75$^{+0.06}_{-0.07}$, $Z$=0.07$^{+0.03}_{-0.02}$  & 25.9 (16) & $kT$=0.63$^{+0.07}_{-0.06}$, $Z$=0.05$^{+0.03}_{-0.02}$ & 27.2 (16) \\[2pt]

& APEC+PL  &  $kT$=0.72$^{+0.06}_{-0.13}$, $Z$=0.13$^{+0.14}_{-0.07}$, $\Gamma$=(1.7)  &  21.2 (15) & $kT$=0.59$^{+0.06}_{-0.10}$, $Z$=0.10$^{+0.12}_{-0.06}$, $\Gamma$=(1.7)  &  20.8 (15)  \\[2pt]

& APEC+PL  &  $kT$=0.74$^{+0.05}_{-0.05}$, $Z$=(1.0), $\Gamma$=(1.7)   &  24.6 (16) & $kT$=0.60$^{+0.04}_{-0.04}$, $Z$=(1.0), $\Gamma$=(1.7) & 24.3 (16)    \\[2pt]

Nucleus & PL+PH(GAU+PL) & $\Gamma_{\rm 1}$=(1.7), $N_{\rm H}$=$24.43^{+2.70}_{-0.56}\times10^{22}$ & 26.0 (26) & &  \\[2pt]
     &             & $E$=$6.32^{+0.02}_{-0.06}$, $\sigma$=(0), $\Gamma_{\rm 2}$=1.34$^{+0.23}_{-0.05}$  &   & $...$ &   $...$ \\[2pt]
     &                 & $EW$ (gau)=0.16$^{+0.06}_{-0.04}$ & &    &    \\[2pt]
     &                 & $L_{\rm bolometric}$ (nuc.)=1.15$^{+0.03}_{-0.05}\times10^{43}$$^g$ &  &  & \\
\hline \hline
\end{tabular}
\vspace{-0.5cm}
\tablenotetext{a} {In this table, we compare parameters and goodness of fits obtained with different models and with two different atomic databases for the X-ray spectral models.}
\tablenotetext{b}{The tail region is the two straight parts (the three green boxes in Figure 4, 35 kpc in length) of the X-ray tail, while the head is the swollen part shown as a green circle (7$''$.87 in radius) excluding the small circle (3$''$ in radius, centered on the X-ray AGN, see Figure 4). The excluded small circle is defined as the nucleus.}
\tablenotetext{c}{The Galactic absorption component (1.73$\times10^{21}$ cm$^{-2}$) is included in all cases.}
\tablenotetext{d}{Obtained with AtomDB v2.0.1.}
\tablenotetext{e}{The units of $kT$, $E$, $\sigma$, and $EW$ are keV; those of $Z$, $L$, and $N_{\rm H}$ are $Z_{\odot}$, erg s$^{-1}$, and cm$^{-2}$, respectively. Parameters in parentheses are fixed.}
\tablenotetext{f}{Obtained with AtomDB v1.3.1.}
\tablenotetext{g}{The luminosity has been corrected for absorption and the bolometric correction adopted is 15 from Vasudevan \& Fabian (2007).}

}
\end{center}
\end{table}

\clearpage

\begin{table}
\begin{center}
{\caption{Pressures in the X-ray Tail and in the Ambient ICM}
\vspace{0.5cm}
\begin{tabular}{cccc} \hline \hline
 Region & \multicolumn{3}{c}{Pressure ($10^{-11}$ dyn cm$^{-2}$)}  \\ \hline

 Tail$^a$ & Single APEC & Double APEC &  CEMEKL \\[1pt]
       & $7.3f^{-1/2}$  & $6.6f^{-1/2}$ & $17.8f^{-1/2}$  \\ \hline
 ICM & Thermal pressure$^b$ & Ram pressure$^b$  & \\[1pt]
 & 4.2 & $15.3(\frac{v_{\rm gal}}{2000 \rm \ km\ s^{-1}})^2$ & \\[2pt]
\hline \hline
\end{tabular}
\vspace{-0.5cm}
\tablenotetext{a}{Pressures are averaged over the full length of the X-ray tail assuming the best-fit abundances. Here $f$ is the volume filling factor.  For the individual regions (Figure 4), the pressures for the single-$kT$ fits range from $4.6f^{-1/2}\times10^{-11}$ dyn cm$^{-2}$ to $9.4f^{-1/2}\times10^{-11}$ dyn cm$^{-2}$, with $\sim30\%$ errors.}
\tablenotetext{b}{Obtained by assuming the cluster-centric distance is the projected distance to the cluster center (see Section 5.1 for a discussion).}
}
\end{center}
\end{table}

\begin{table}
\begin{center}
{\caption{\ga\ vs. \gab }
\vspace{0.5cm}
\begin{tabular}{llcc} \hline \hline
 Region & Parameter & \ga $^a$  & \gab $^a$  \\ \hline 
 Tail & $l \times w$ (kpc $\times$ kpc)$^b$ & $\sim80\times6$, $80\times4$ & $\sim40\times3$ \\
 & $M_{\rm tail}$ ($10^8 M_{\rm \odot}$)$^{c,d}$ & $12.9f^{1/2}$ ($13.8f^{1/2})$ & $2.0f^{1/2}$ ($2.3f^{1/2}$) \\
 & $f_{\rm X-ray}$$^e$ & $25.8\%f^{1/2}-43.0\%f^{1/2}$ & $2.2\%f^{1/2}-5.0\%f^{1/2}$ \\
 & $L_{\rm 0.5-2\ keV}$(tail) (10$^{40}$ erg s$^{-1}$)$^{d,f}$ & 8.2 (8.3) & 3.1 (3.2) \\
 & $L_{\rm bolometric}$(tail) (10$^{40}$ erg s$^{-1}$)$^{d,f}$ & 17 (18)  & 6.0 (6.6) \\
 & $kT_{\rm tail}$ (keV)$^d$ & 0.92$\pm0.04$ (0.81$\pm$0.03) & 0.95$^{+0.04}_{-0.05}$ (0.80$^{+0.03}_{-0.04}$) \\
 & $n_{\rm tail}$ (cm$^{-3}$)$^{d,g}$ & 0.014$f^{-1/2}$ (0.015$f^{-1/2}$)  & 0.025$f^{-1/2}$ (0.029$f^{-1/2}$) \\
 ICM & $kT_{\rm ICM}$ (keV) & $\sim6.5$ (6.3) & $\sim7.1$ (7.0) \\
 & $n_{\rm ICM}$ (10$^{-3}$ cm$^{-3}$) & 1.4 & 1.9 \\
 & $P_{\rm ram}$ (10$^{-11}$ dyn cm$^{-2}$) & 11.6$(\frac{v_{\rm gal}}{2000\ {\rm km\ s^{-1}}})^2$ & 15.3$(\frac{v_{\rm gal}}{2000\ {\rm km\ s^{-1}}})^2$ \\
 & $P_{\rm thermal}$ (10$^{-11} $dyn cm$^{-2}$) & 2.9 & 4.2 \\
 Nucleus & $L_{\rm 0.3-10\ keV}$(nuc.) (erg s$^{-1}$)$^h$ & $<1.2\times10^{39}$ & 1.0$\times10^{42}$ \\
\hline \hline
\end{tabular}
\vspace{-0.5cm}
\tablenotetext{a}{Some of the values assume the cluster-centric distances are the projected distances (see Sections 5.1 and 5.3). For results containing parentheses (except the row of $P_{\rm ram}$), the outside ones were obtained with AtomDB v2.0.1 while the inside ones were obtained with AtomDB v1.3.1.}
\tablenotetext{b}{Size of the X-ray tail. The two data sets for \ga 's tails are the sizes of the main and secondary tails. Length refers to the full length of the tail (including the ``head''), while width refers to the FWHM, so the widths of \ga 's tails are different from those in S10.}
\tablenotetext{c}{Mass of the X-ray tail(s). The volume of the tail is assumed to be a cylinder with the size given above. $f$ is the filling factor of the soft X-ray emitting gas. Note that the mass of \ga 's tails has been rescaled according to their new volume compared to the old one in S10.}
\tablenotetext{d}{Obtained from the single-$kT$ fit.}
\tablenotetext{e}{Mass ratio of the X-ray tail(s) to the total ISM of the galaxy from S10.}
\tablenotetext{f}{Intrinsic luminosity of the X-ray tail(s). The error of the luminosity is $\sim5\%$ for \ga\ and $\sim4\%$ for \gab .}
\tablenotetext{g}{Average electron density of the X-ray tail(s). The values for \ga 's tails have been rescaled.}
\tablenotetext{h}{Assuming no intrinsic absorption for \ga 's nucleus, while absorption for \gab 's nucleus is adopted from the best-fit value, i.e., 2.44$\times10^{23}$ cm$^{-2}$. The error of \gab 's nuclear luminosity is $\sim4\%$.}
}
\end{center}
\end{table}

\clearpage

\begin{figure}
\centerline{\includegraphics[height=0.6\linewidth]{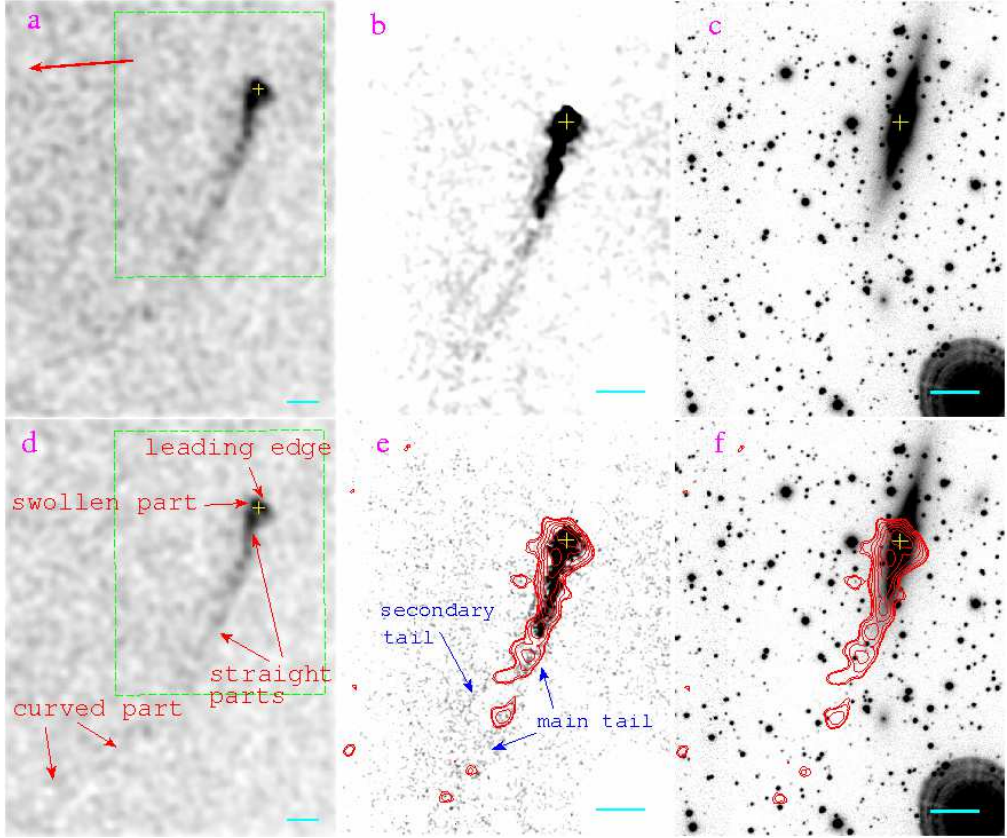}}
\vspace{0.5cm}
\scriptsize
  \caption{{\bf a}: \chandra\ $0.5-2.0$ keV count image of \gab . No background subtraction or exposure correction was applied. Point sources were removed and the image was smoothed with a 3$''$.94 Gaussian kernel. The green dashed box is the field of view (FOV) of the SOAR images shown on the right. The red arrow denotes the direction to the X-ray peak of the cluster. A long X-ray tail is clearly seen from the image.
{\bf b}: Net H$\alpha$ image of \gab\ from SOAR. The image was background subtracted and smoothed with a 1$''$.5 Gaussian. Stars were masked. A significant H$\alpha$ tail is visible, with two branches bifurcating at the far end. 
{\bf c}: Optical continuum image of \gab\ from SOAR. This image has the same size as panel \textbf{b}. The Galactic foreground stars are crowded in the FOV.
{\bf d}: \chandra\ $0.5-2.0$ keV image of \gab\ with background subtraction and exposure correction. The image was smoothed with a 3$''$.94 Gaussian. The long X-ray tail can further be subdivided into four parts based on morphology (see the text for details), as indicated in the figure. At the upstream side of the tail, a sharp leading edge is also present. At the position of the secondary H$\alpha$ tail (panel \textbf{e}), X-ray emission is enhanced at the $\sim2\sigma$ level.
{\bf e}:  The $0.5-2.0$ keV \chandra\ contours in red superposed on the net H$\alpha$ image. The H$\alpha$ image was smoothed with a 0$''$.6 Gaussian. The two H$\alpha$ tails are marked, while the main tail spatially coincides with the X-ray tail. 
{\bf f}: The optical continuum image of \gab\ with the $0.5-2.0$ keV \chandra\ contours in red overlaid. In all panels, the location of the X-ray AGN is marked with a yellow cross, while the cyan scale bars are 5 kpc (or 15$''$.28).
}
\end{figure}

\begin{figure}
\centerline{\includegraphics[height=0.6\linewidth]{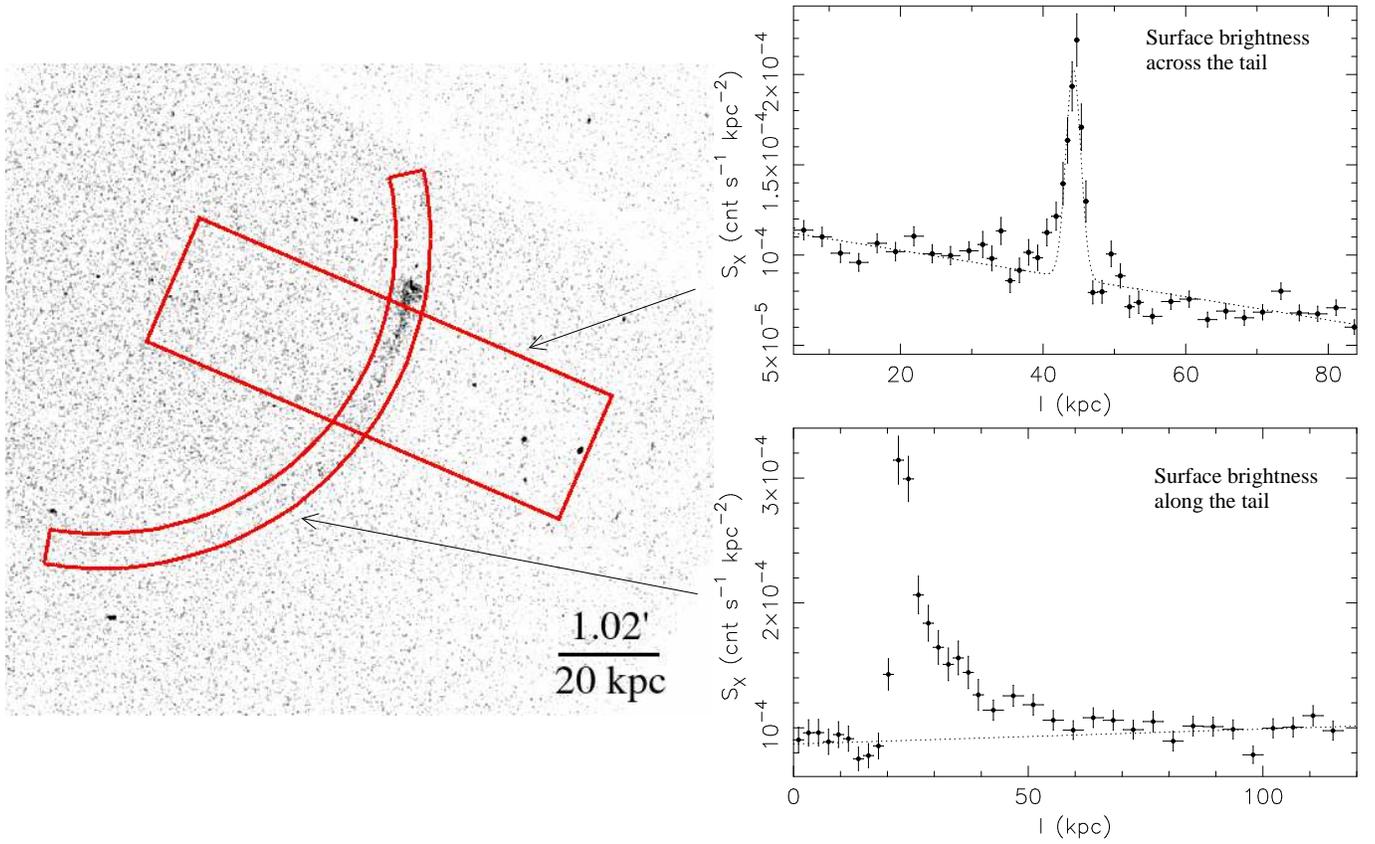}}
\vspace{0.5cm}
  \caption{The \chandra\ $0.6-2.0$ keV surface brightness profiles across and along the X-ray tail of ESO 137-002. The regions used to measure the surface brightness are shown on the left. The widths of the regions are 80\arcsec\ and 20$''$.7, respectively, with the physical and angular scales shown in the lower right corner. The dotted line in the upper right panel represents the local background plus a Gaussian which is used to model the emission from the tail, while that in the lower right panel represents the local background. Note that the background around the tail is high, so the jump across the edge is relatively small (a factor of $\sim3$). A Gaussian fit gives a FWHM of 2.9$\pm$0.2 kpc for the width of the tail. The tail is significant to $\sim40$ kpc from the nucleus, with a possible extension to large distances. The flux conversion factor for the tail is: 10$^{-4}$ cts s$^{-1}$ = 4.85$\times10^{38}$ erg s$^{-1}$ (bolometric).
}
\end{figure}
\clearpage

\begin{figure}
\centerline{\includegraphics[height=0.4\linewidth]{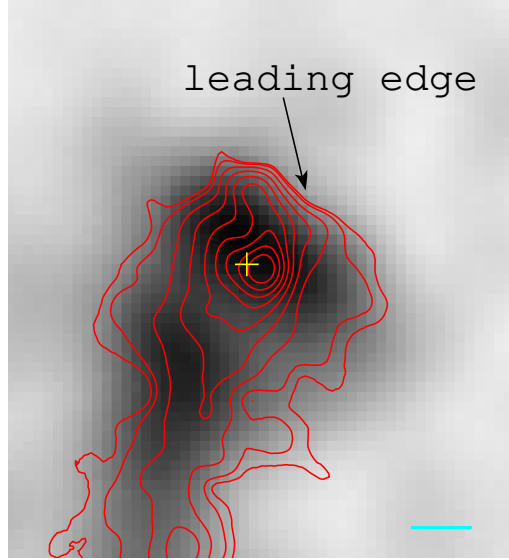}}
\vspace{0.5cm}
  \caption{\chandra\ $0.5-2.0$ keV image of the nuclear region with the net H$\alpha$ contours overlaid in red. The X-ray leading edge is spatially coincident with the H$\alpha$ edge. The location of the X-ray AGN is marked with a yellow cross. The offset between the nucleus and the H$\alpha$ peak is probably caused by the dust lane in the bulge. The cyan scale bar is 1 kpc.
}
\end{figure}

\begin{figure}
\centerline{\includegraphics[height=0.5\linewidth]{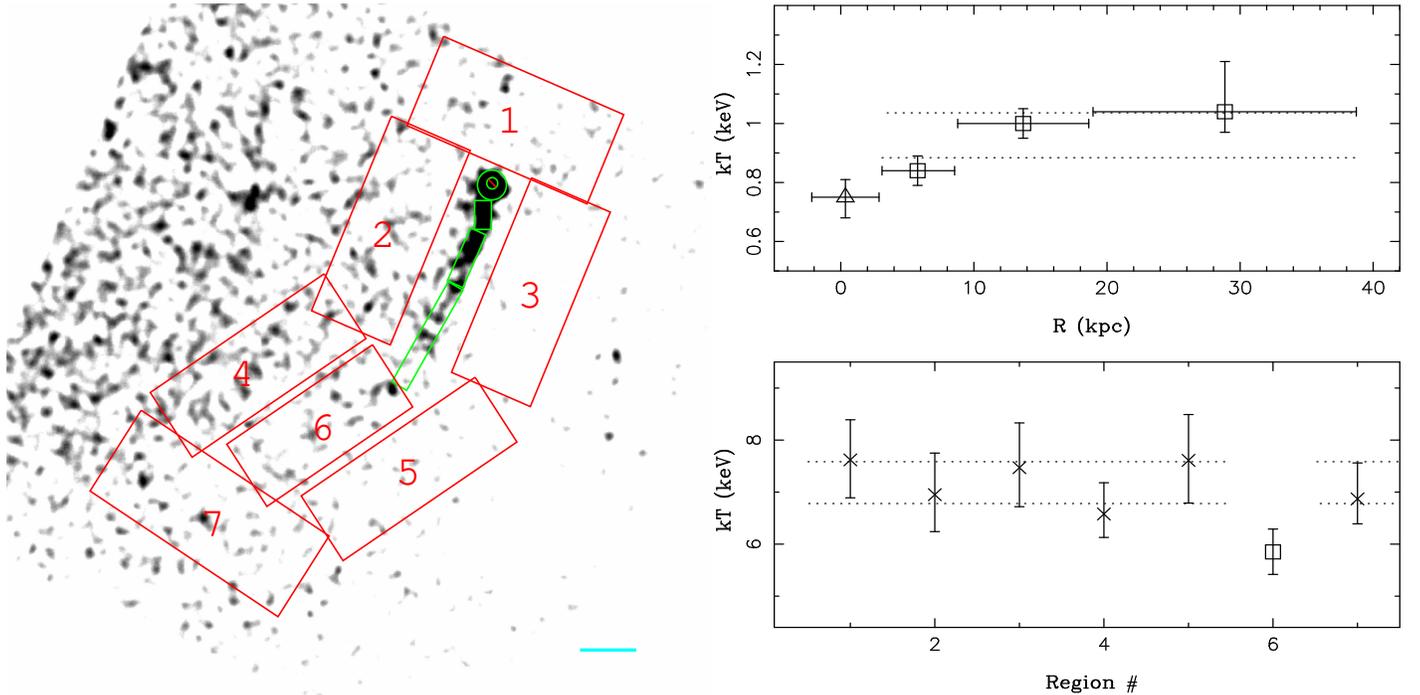}}
\vspace{0.5cm}
  \caption{{\bf Left}: Regions where the temperatures of the X-ray tail and the ICM are measured. The tail regions are in green while the ICM regions are in red numbered from 1 to 7. The ``head'' (swollen part) is the green circle, with the small circle (\textbf{3$''$} in radius, centered on the X-ray AGN) excluded. The small circle excluded from the ``head'' is defined as the nucleus. Both the ``head'' and the nucleus were excluded from the global spectral fits. Point sources were removed. The cyan scale bar is 10 kpc. {\bf Right}: Temperatures (with $1\sigma$ error bars) for the tail (upper) and the ICM (lower) regions shown on the left. In the upper panel, the temperature of the ``head'' is shown as a triangle with those of the other three regions shown as squares. The horizontal error bars are the bin sizes of the individual regions along the tail. The position of the X-ray AGN is at the origin. The temperatures of the outer three regions were obtained by linking the abundances of these three regions, so the actual temperature uncertainties should be larger than those shown. The two dotted lines represent the $1\sigma$ range of the temperatures for the total spectrum of the outer three regions. The temperature variations are within $1.4\sigma$, so the whole tail has a nearly constant temperature. In the lower panel, the dotted lines represent the $1\sigma$ temperature range of the 6 regions excluding region \# 6. Region \# 6 has the lowest temperature, which is likely a hint of the contamination by stripped gas there. A similar ICM temperature drop behind the observed tail is also found for \ga\ (S10).
}
\end{figure}

\begin{figure}
\centerline{\includegraphics[height=0.5\linewidth]{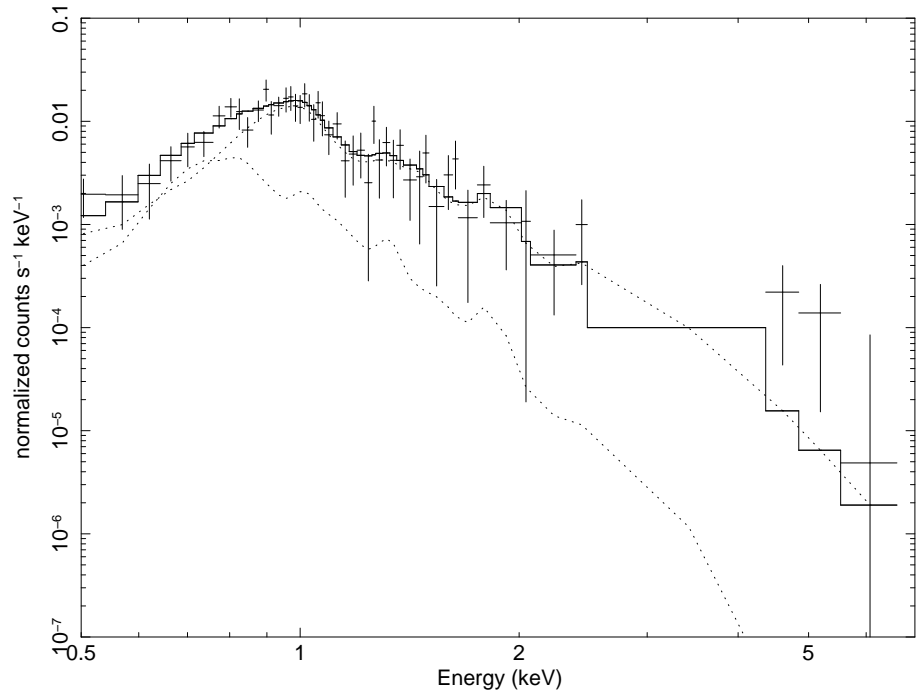}}
\vspace{0.5cm}
  \caption{X-ray spectrum of the global tail region (excluding the ``head''). The double-$kT$ model fit is also shown as a solid line, with the two dotted lines representing the two thermal model components. 
}
\end{figure}

\begin{figure}
\centerline{\includegraphics[height=0.5\linewidth]{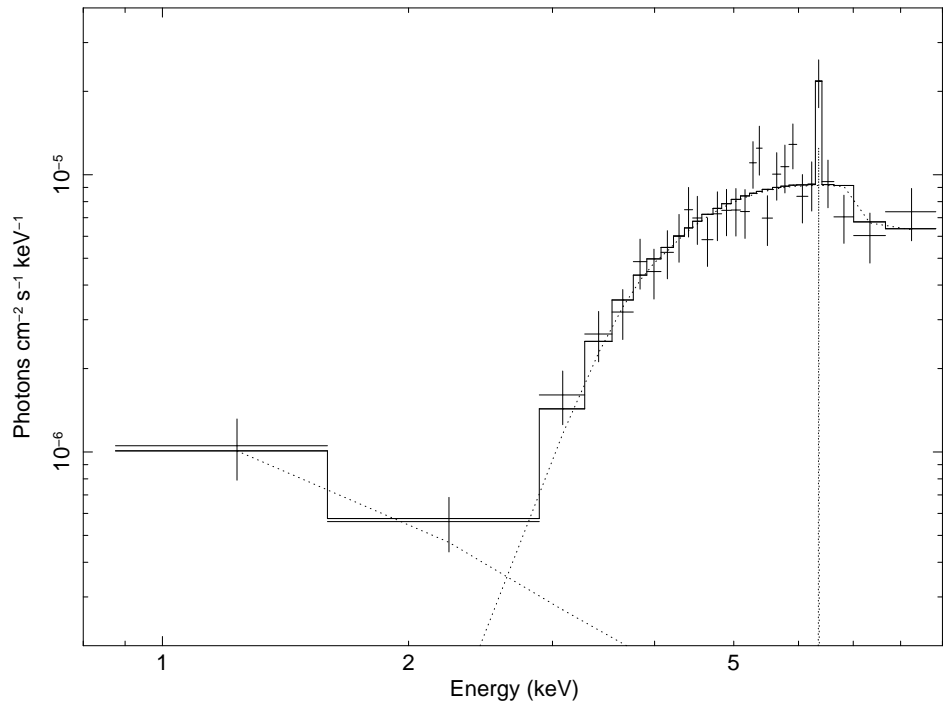}}
\vspace{0.5cm}
  \caption{Unfolded X-ray spectrum (see the text) of the nucleus. The model is also shown as a solid line, while the three dotted lines are the three model components including a redshifted iron K$\alpha$ fluorescence line.
}
\end{figure}

\end{document}